\documentclass[conference]{IEEEtran}
\IEEEoverridecommandlockouts
\usepackage{cite}
\usepackage{amsmath,amssymb,amsfonts}
\usepackage{algorithmic}
\usepackage{graphicx}
\usepackage{textcomp}
\usepackage{xcolor}

\usepackage{multirow}
\usepackage[hyphens]{url}
\usepackage{verbatim}
\usepackage{tabularx}
\usepackage{booktabs}
\usepackage{comment}
\usepackage{enumitem}

\usepackage{tikz}

\newcommand\copyrighttext{%
  \footnotesize \textcopyright \the\year{} IEEE. Personal use of this material is permitted. Permission from IEEE must be obtained for all other uses, including reprinting/republishing this material for advertising or promotional purposes, collecting new collected works for resale or redistribution to servers or lists, or reuse of any copyrighted component of this work in other works.}

\newcommand\copyrightnotice{%
\begin{tikzpicture}[remember picture,overlay]
\node[anchor=south,yshift=10pt] at (current page.south) {\fbox{\parbox{\dimexpr0.75\textwidth-\fboxsep-\fboxrule\relax}{\copyrighttext}}};
\end{tikzpicture}%
}

\def\BibTeX{{\rm B\kern-.05em{\sc i\kern-.025em b}\kern-.08em
    T\kern-.1667em\lower.7ex\hbox{E}\kern-.125emX}}
\begin{document}

\title{Where are the Frontlines? A Visualization Approach for Map Control in Team-Based Games\thanks{Short Paper}}

\author{\IEEEauthorblockN{Jonas Pech\'e}
    \IEEEauthorblockA{\textit{Computer Graphics} \\
    \textit{Johannes Kepler University Linz}\\
    Linz, Austria \\
    j\_peche@wargaming.net}
\and
    \IEEEauthorblockN{Aliaksei Tsishurou}
    \IEEEauthorblockA{\textit{Independent Researcher} \\
    Wroclaw, Poland \\
    aliakseitsishurou@gmail.com}
\and    
    \IEEEauthorblockN{Alexander Zap}
    \IEEEauthorblockA{\textit{DS Research} \\
    \textit{Wargaming}\\
    Berlin, Germany \\
    a\_zap@wargaming.net}
\and
    \IEEEauthorblockN{G\"unter Wallner}
    \IEEEauthorblockA{\textit{Computer Graphics} \\
    \textit{Johannes Kepler University Linz}\\
    Linz, Austria \\
    guenter.wallner@jku.at}
}

\maketitle
\copyrightnotice

\begin{abstract}
    A central area of interest in many competitive online games is spatial behavior which due to its complexity can be difficult to visualize. Such behaviors of interest include not only overall movement patterns but also being able to understand which player or team is exerting control over an area to inform decision-making. Map control can, however, be challenging to quantify. In this paper, we propose a method for calculating frontlines and first efforts towards a visualization of them. The visualization can show map control and frontlines at a specific time point or changes of these over time. For this purpose, it utilizes support vector machines to derive frontlines from unit positions. We illustrate our algorithm and visualization with examples based on the team-based online game \emph{World of Tanks}. 
\end{abstract}


\section{Introduction}
    \label{sec:introduction}

Visualizations depicting various facets of gameplay have long been employed in the context of games user research and analytics to better understand player behavior, to unearth design issues, and to inform game development. More recently, visualizations of game data have also gained traction within various player communities to assist in gameplay review and skill development. These can be used in -- what Medler~\cite{medler2011playerdossiers} called -- player dossiers that provide feedback after play directly within a game, external websites, and in tools for post-reflection or live feedback. For instance, commercial platforms such as \emph{Shadow.GG}~\cite{shadowgg2017}, \emph{STRATZ}~\cite{stratz2024}, and \emph{Leetify}~\cite{leetify2025} provide post-game insights through summaries, depiction of movement patterns, and different types of heatmaps. Some games even integrate extensive analytics and visualizations directly in the client such as the team-based multiplayer online battle arena (MOBA) game \emph{League of Legends}~\cite{leagueoflegends}.

Such training tools and player-facing visualizations have also found increasing scholarly interest (e.g.,~\cite{afonso2019visualeague,philip2020,ma2025stratincon}). Often such visualizations directly present the unprocessed individual data points (e.g., death locations). However, Wallner et al.~\cite{Wallner:2021} -- studying players' information needs with respect to post-play visualizations for three competitive game genres (MOBA, real-time strategy, battle royale) -- found that while players are being interested in such low-level data, they also have a strong interest in being able to observe higher-level information derived from the low-level data.

One key area of interest in many competitive online games is spatial behavior which, due to its complexity, benefits from aggregated representations. For instance, Wallner~\cite{Wallner:2018,Wallner:2019} used the concept of battle maps to summarize troop movements and the overall flow of a battle. Kuan et al.~\cite{kuan2017starcraft} developed similar visualizations of battle dynamics for \emph{StarCraft II} to show unit movements, attack lines, and zones of interest. However, spatial aspects other than the visualization of overall movement patterns have received less attention.

One such aspect is map control, which in the study of Wallner et al.~\cite{Wallner:2021} was deemed highly relevant for MOBA and real-time strategy games. Among others, it allows players to restrict the movement of the opposing team, to gain information about what it is doing, and to influence their decision-making (e.g.,~\cite{web:mapcontrol1,web:mapcontrol2}). Thus, map control deals with the understanding of who is controlling which parts of the map and how the frontline -- the separation between two or more opposing teams -- changes over time. However, while map control itself is a crucial part of many games, it is difficult to quantify with currently limited work on finding frontlines within team-based games. This, in turn, hampers visualization efforts. For instance, while Ma et al.\cite{ma2025stratincon} -- proposing the \emph{StratIncon Detector}, a visual analytics tool for MOBAs that identifies deviations from the ideal strategic behavior to improve the outcome of battles -- touched upon the matter of map control, they do not explicitly visualize it. In the context of artificial intelligence for games, influence maps are frequently used to map out enemy territory~\cite{web:influencemaps} but these rather serve as internal representation although they could be visualized as heatmaps. Nevertheless, heatmaps do not provide an explicit representation of the frontline.

\begin{figure*}[t]
    \centering
    \includegraphics[width=1.0\linewidth,trim={0cm 0.75cm 0cm 0.75cm},clip]{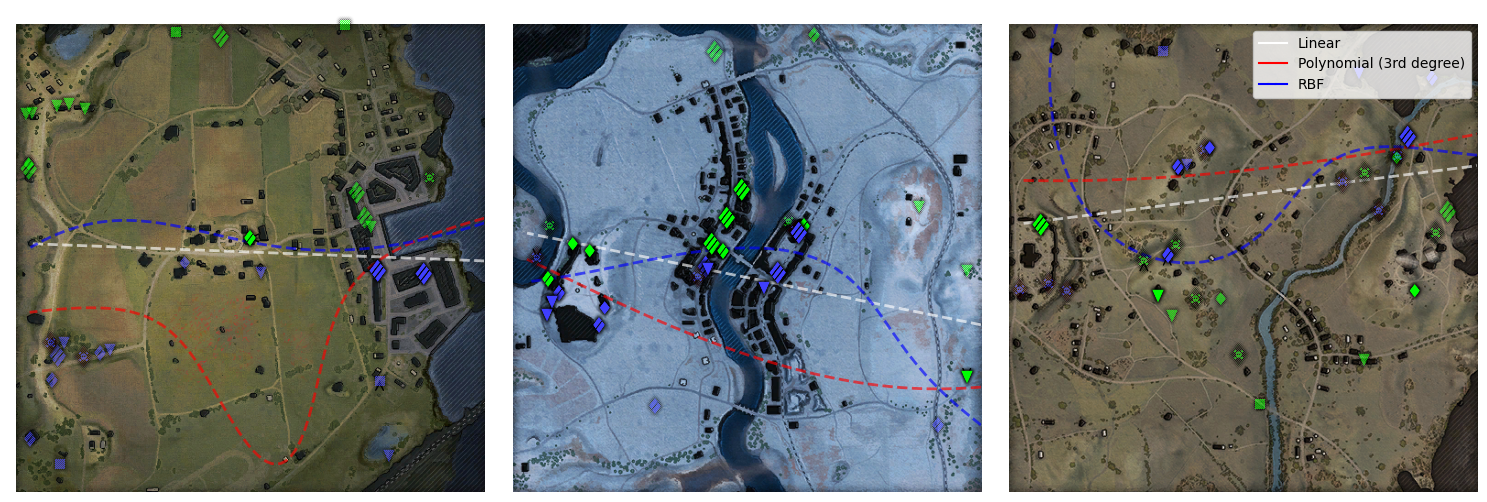}
    \caption{The RBF, polynomial, and linear baseline kernel visualized over four situations in different states of the game. Vehicles are visualized in their team color (blue and green) with different icons indicating different vehicle types. The linear kernel is shown in white, the polynomial kernel of degree three in red, and the RBF kernel in blue. The linear kernel represents the frontline insufficiently while the suboptimal polynomial line tends to overshoot.}
    \label{fig:kernels}
\end{figure*}

In this paper, we propose an algorithm and visualization for conveying map control, particularly frontlines in team-based games. The visualization allows observing changes in frontlines over time in single static images as well as through overlaid frames forming a ghosting effect. We demonstrate our approach with data from \emph{World of Tanks} (\emph{WoT})~\cite{game:wot}, a team-based online multiplayer game in which two teams of players (each player controlling a single tank) compete against each other for different map objectives. Map control is an essential part of its gameplay, providing high-level information about the battle's development and the teams' strategies, making it an appropriate use case for demonstrating the approach. However, the presented approach can be applied to similar team-based games as well.

\section{Approach}
    \label{sec:approach}

Given our team-based context, we define a \emph{frontline} as the separation between two sets of vehicles, one for each team. The line should be as simple as possible, while still maintaining a reasonable margin separating the vehicles. Furthermore, situations which are not clearly separable can be quite common, therefore, the frontline must be stable against outliers. Typical solutions for this type of problem are support vector machines (SVMs)~\cite{svm} as defined in Equation~\ref{eq:svm}. 
\begin{equation}
    \begin{aligned}
    \max_{\alpha} \quad & \sum_{i=1}^n \alpha_i - \frac{1}{2} \sum_{i=1}^n \sum_{j=1}^n \alpha_i \alpha_j y_i y_j K(x_i, x_j) \\
    \text{subject to} \quad & 0 \leq \alpha_i \leq C, \quad \sum_{i=1}^n \alpha_i y_i = 0
    \end{aligned}
    \label{eq:svm}
\end{equation}

\noindent SVMs find the hyperplane that maximizes the margin between support vectors of the two sets.\footnote{While generally binary, an SVM can also be applied to multi-class problems (e.g., games with three teams) by treating the labels as a One-vs-Rest -- a strategy that trains one classifier per class, treating all other classes as a single negative class.} This is achieved by solving a constrained optimization problem using Lagrange multipliers. However, a linear front rarely makes sense in most game scenarios; therefore, an appropriate kernel is required to map the vehicle positions to a higher-dimensional, nonlinear space that can be separated by a hyperplane more easily. Generally, two common kernels are available:

\begin{itemize}[leftmargin=*]
    \item \textbf{RBF}: A radial basis kernel that measures similarity based on distance defined as $K(x, x') = \exp\left(-\gamma \|x - x'\|^2\right)$
    \item \textbf{Polynomial}: Computes similarity as a polynomial function.
\end{itemize}

\noindent Polynomial kernels are, however, not the best choice for frontlines as well, since they require more tuning and have an additional \emph{degree} hyperparameter. The fitted margin of a polynomial kernel is also less consistent and visually appealing than the one of an RBF kernel, as illustrated in Figure~\ref{fig:kernels}. The figure shows the difference between the different kernels and highlights the unexpected fitted lines of the polynomial kernel. For the polynomial kernel, we picked a degree of three based on visual inspection.

\begin{figure*}
    \centering
    \includegraphics[width=1.0\linewidth,trim={0cm 1.5cm 0cm 0.6cm},clip]{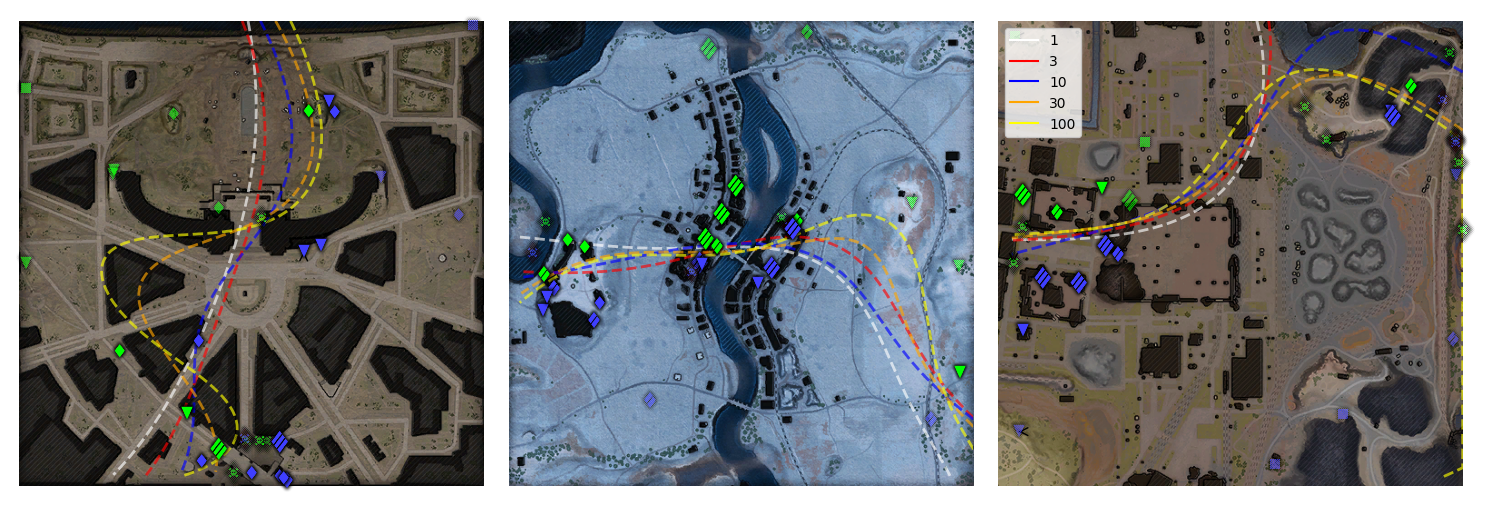}
    \caption{Different values for different regularization parameters \textit{C}. Higher values for \textit{C} result in less regularization and therefore in more pronounced frontlines.}
    \label{fig:c}
\end{figure*}

Other parameters are the gamma parameter, which we set to $1 / n_{\text{features}} \cdot \operatorname{Var}(X)$ to obtain well-defined and scaled values independent from the map dimensions. In our use case we set $n_{\text{features}}$ to 2 since we treat situations as 2D positions. Parameter \textit{C} affects how outliers, or vehicles breaching the frontline, impact the frontline. Mathematically, it defines the trade-off between maximizing the margin and minimizing classification error. This value needs to be tuned more carefully as it controls the soft threshold of frontline breaks and how strong outliers bend the frontline in their favor. Figure~\ref{fig:c} shows the effect of different \textit{C} on an approximately logarithmic scale in different scenarios. One can see how the frontline fluctuates more in complex scenarios with high \textit{C} values, especially visible in the left figure. Based on our specific focus on \textit{WoT} battles, we empirically picked, through manual experimentation, a C of 10 for all further visualizations presented here.

\begin{figure*}
    \centering
    \includegraphics[width=0.99\linewidth,trim={0cm 1.5cm 0cm 0.6cm},clip]{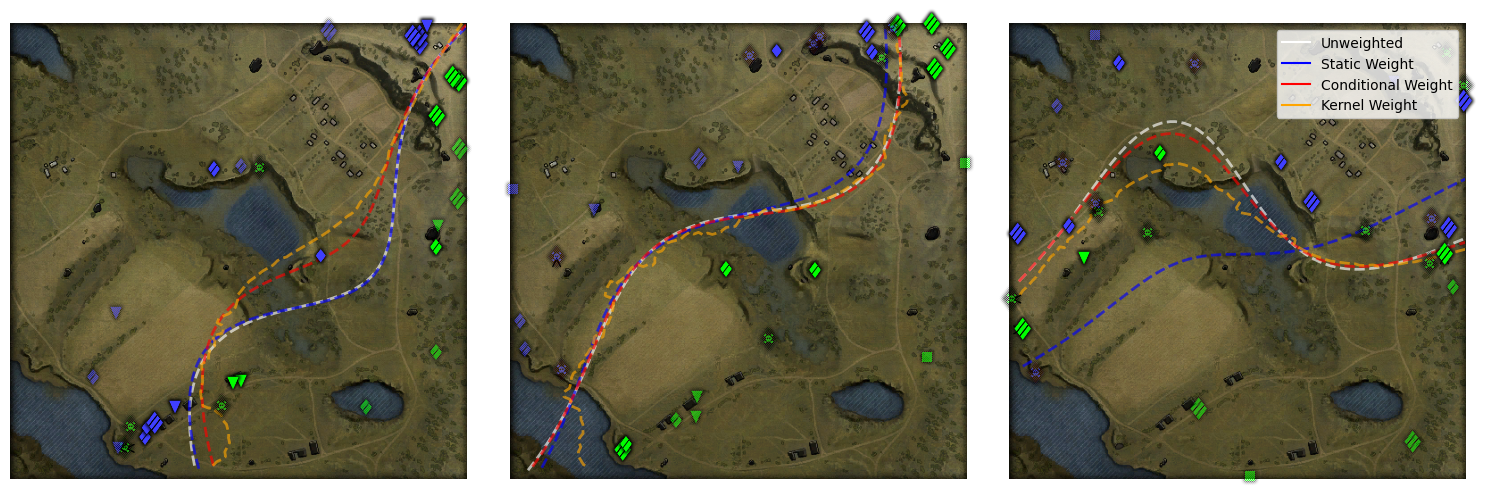}
    \caption{Different weighting approaches affect the frontlines, shifting the margin towards vehicles with lower weight. The white line represents no weighting, blue shows a static weight based on health, red a conditional weight based on the average speed, and orange the same weight as red but applied as a kernel weight.}
    \label{fig:weight}
\end{figure*}

\begin{figure*}
    \begin{minipage}[t]{0.36\linewidth}
            \centering
            \includegraphics[width=1.0\linewidth]{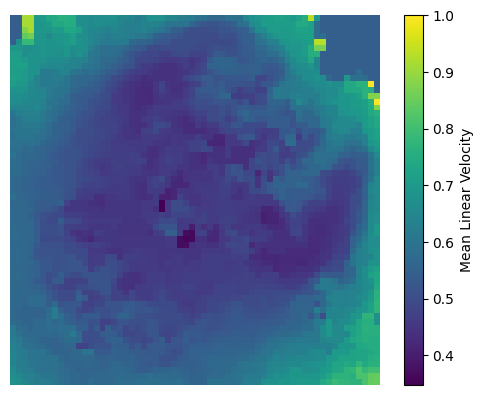}
            \caption{An example weight lookup map, in this case representing the average speed of advance of vehicles.}
            \label{fig:weights}
    \end{minipage}
    \hfill
    \begin{minipage}[t]{0.29\linewidth}
        \centering
        \includegraphics[width=1.0\linewidth]{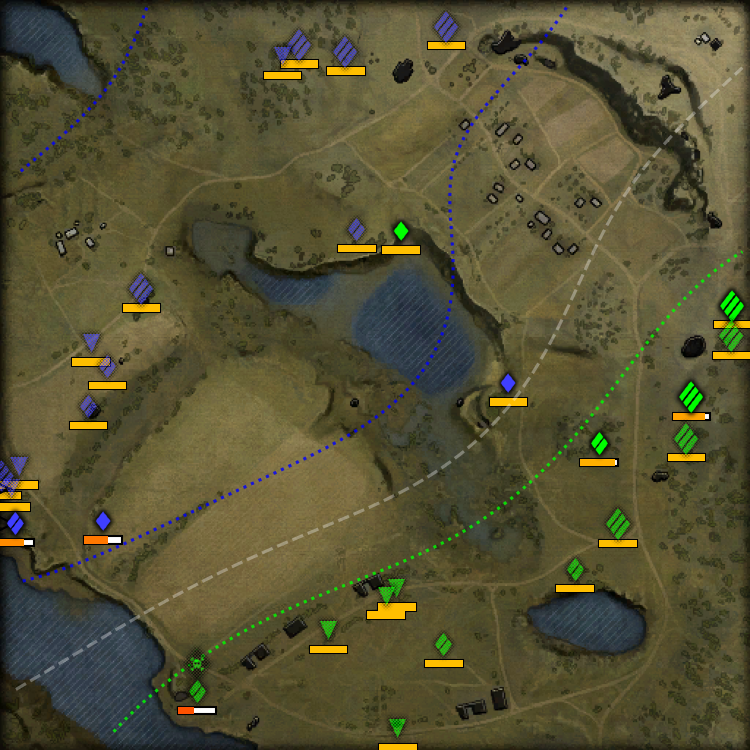}
        \caption{The center frontline in dashed, and 0.75 level frontlines for each team visualized as dotted lines.}
        \label{fig:offset}
    \end{minipage}
    \hfill
    \begin{minipage}[t]{0.29\linewidth}
        \centering
        \includegraphics[width=1.0\linewidth]{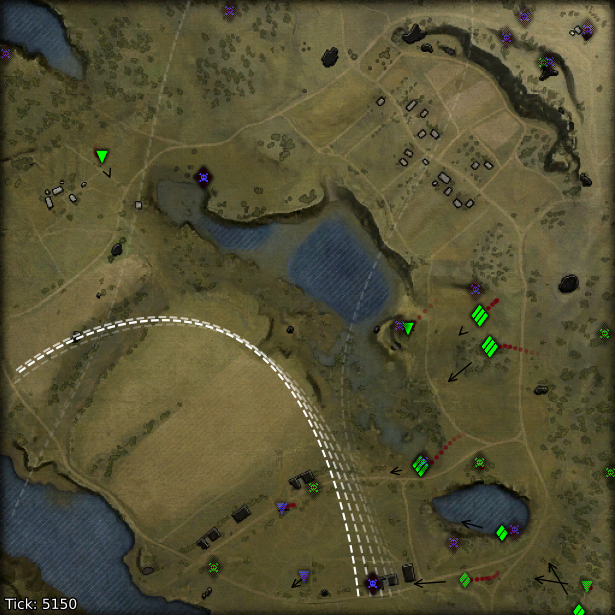}
        \caption{The development of the frontline the moment the area fraction changed the most, visualized as ghost lines in 2 second intervals.}
        \label{fig:break}
    \end{minipage}
\end{figure*}

On top of the kernel, the weight per vehicle can be provided to define the relevance of vehicles aside from their spatial distance to each other. Internally, the weight is multiplied by \textit{C}. Those can be independent factors such as the fraction of remaining health, the players' skill, or different vehicle statistics. Likely more interesting are, however, vehicle dependencies such as visibility, effort for a vehicle to claim another vehicle’s area, or tactical benefit over the enemies' location. By adding such values, the frontline is better fitted on metrics that matter to the overall game. Figure~\ref{fig:weight} shows the unweighted line, a line weighted by the fraction of remaining health, and a third one using potential map advancements (a per-map estimation of velocity with which the vehicles tend to advance from that position, shown in Figure~\ref{fig:weights}).

Additionally, instead of simple weighting, a custom kernel function can be used to control the relation between the samples directly. For example, we can multiple the pairwise weight of two samples $w(x, x')$ with the exponent, yielding:
\begin{equation}
    K(x, x') = \exp\left(-\gamma \, w(x, x') \, \|x - x'\|^2\right)
\end{equation}
This effectively transforms the distance between two vehicles by a custom weight based on map characteristics, battle dynamics, or similar available data. This approach, however, needs careful tuning and design of the weight function \emph{w} since an inconsistent or non-continuous function will result in a non-continuous frontline.


To visualize the obtained center frontline, we use the decision boundary (the points in space where the fitted decision function, as defined in Equation~\ref{eq:svm_decision_function}, evaluates to $f(x) = 0$) and also the frontline of each team at different levels ($f(x) = \text{level}$) as shown in Figure~\ref{fig:offset}. 
\begin{equation}
    f(x) = \sum_{i \in \text{SV}} \alpha_i y_i K(x_i, x) + b
    \label{eq:svm_decision_function}
\end{equation}

\noindent A level of -1 and 1 is the margin on which the support vectors are located. The exact value depends on the use case: for example, we picked $\pm 0.75$ to represent the approximate start of the area under conquest, since the actual frontline should be in front of the vehicles and not at their position. Or, for example, a value of $\pm 1.25$ can be used to get the approximate line for light vehicles, behind the line of heavy vehicles (an important classification for vehicles in \emph{WoT}).

\noindent One drawback of the RBF kernel is the extension of frontlines at specific levels behind the respective teams. While mathematically correct, this results in a misleading visualization. One solution is to clip or fade the line if it gets too far from the center frontline, or when the gradient of the decision function points away from the center of action. In Figure~\ref{fig:offset} this small secondary blue line, which should be removed, appears in the upper left corner.

\begin{figure}[b]
    \centering
    \includegraphics[width=0.9\linewidth]{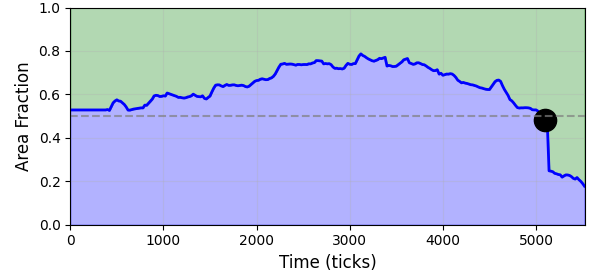}
    \caption{The fraction of the occupied area for both teams, with the point of highest deviation $\bullet$ highlighted.}
    \label{fig:area}
\end{figure}

Frontline computation also enables the measurement of further metrics such as the area occupied by a team. This can be a useful metric for measuring positive or negative frontline development. The area is defined as the fraction of positions \textit{x} for which $f(x) > 0$. Figure~\ref{fig:area} shows the fraction of controlled area over time, clearly visualizing the moment Team 1 (green) started to win. At this time point, the blue team lost two tanks, and another one is retreating, pulling back the frontline. This moment is also visualized in Figure~\ref{fig:break} in the form of ghost lines in 2-second intervals. Moreover, the raw decision function itself can be treated as an influence map, giving rise to further use cases.

Optimized SVM implementations allow this method to run in under 2~ms on an i9-13900HX CPU, single-threaded, and with 1024 samples for area control estimation. More units will increase runtime while constraining fitting iterations and sampling resolution can reduce it.

\section{Conclusions and Future Work}
    \label{sec:conclusions}

This paper showcased how SVMs can be used to calculate frontlines in team-based combat games, and how to customize the frontline estimation using vehicle weights, decision function levels, and hyperparameter adjustments. We also highlighted potential uses beyond visualization such as measuring the area occupied by each team. Several directions for future work exist such as the mentioned improvement of back-facing frontlines as seen in Figure~\ref{fig:offset}, or a more sophisticated weighting and kernel functions to better adapt to map and game dynamics. The area estimation, only briefly mentioned, can also be expanded on to weight the area by strategic value, differentiate the area weight between captured and contested areas, or consider other custom and game-specific weighting factors. One limitation is having only a single frontline. While the RBF kernel can efficiently define a complex frontline, if the team splits up into multiple fronts, the visualization becomes hard to interpret. There are various potential fixes to improve on this issue such as introducing additional classes and letting the natural multi-class design of SVM handle that.

\section*{Acknowledgments}

We would like to thank \emph{Wargaming} for providing a high-quality dataset to experiment and evaluate on.

\bibliographystyle{IEEEtran}
\bibliography{references}

\end{document}